\documentclass[apjl]{emulateapj}
\usepackage{epsfig}
\begin{document}

\title{A new mechanism for radial migration in galactic disks: spiral-bar resonance overlap}

\author{I.~Minchev\altaffilmark{1} \& B.~Famaey\altaffilmark{1}}

\altaffiltext{1}{Universit\'e de Strasbourg, CNRS, Observatoire Astronomique, 11 rue
de l'Universit\'e, 67000 Strasbourg, France; ivan.minchev@astro.unistra.fr}

\begin{abstract}
While it has long been known that a large number of short-lived transient spirals can 
cause stellar migration, here we report that another mechanism is also effective at 
mixing disks of barred galaxies. The resonance overlap of the bar and spiral structure
induces a nonlinear response leading to a strong redistribution of angular momentum in 
the disk. We find that, depending on the amplitudes of the perturbers, the changes
in angular momentum, $\Delta L$, could occur up to an order of magnitude faster than in 
the case of recurrent spirals. The signature of this mechanism is a bimodality in $\Delta L$
with maxima near the bar's corotation and its outer Lindblad resonance; this is 
independent of the properties of the spiral structure. For parameters 
consistent with the Milky Way the disk mixes in about 3~Gyr and the stellar
velocity dispersion increases with time in a manner roughly consistent with observations. 
This new mechanism could account for both the observed age-velocity relation and the 
absence of age-metallicity relation in the solar neighborhood.
Spiral-bar interaction could also explain observations showing that strongly barred
galaxies have weaker metallicity gradients than weakly barred or non-barred
galaxies.
\end{abstract}


\section{Introduction}

The metallicity of the interstellar medium (ISM) in a galaxy is expected to increase with 
time due to the progressive enrichment through stellar feedback. Consequently, one would 
expect that younger stars formed at the same galactic radius would have higher 
metallicities. In addition, the ISM metallicity has been well established to decrease 
with increasing galactic radius, $r$, due to a more efficient star formation and enrichment 
of the ISM in the central regions of galaxies \citep{wilson94,daflon04}. This in turn would 
result in coeval stars being progressively more metal-poor as $r$ increases. 

However, the metallicities of stars in the solar neighborhood (SN) have been observed to 
lack the expected correlation with age \citep{edvardson93,haywood08}. Re-examining the 
age-metallicity distribution of the Geneva-Copenhagen survey \citep{nordstrom04}, 
\cite{haywood08} found that the data are only consistent with the age-metallicity relation 
(AMR) for stars younger than 3~Gyr. For other samples the width of the metallicity 
distribution was found to be larger than what could be accounted for by measurement errors 
and/or inhomogeneous ISM. Because of the age-velocity relation (AVR), where older stellar 
samples are observed to have higher velocity dispersion than younger ones (e.g. 
\citealt{holmberg09}), stars from the inner and outer Milky Way (MW) disk can enter the SN and 
thus blur the expected AMR. However, this can only account for about 50 \% of the observed 
scatter \citep{nordstrom04}. 

A possible explanation for this discrepancy is the radial migration of stars from their 
birth guiding radii \citep{haywood08,schonrich09}. Indeed, there is now considerable 
evidence that the non-axisymmetry of the Galactic potential can cause significant 
perturbations in the motion of both stars and gas: observational evidence for this comes 
from (i) the observed non-circular motions of gas flows in the inner MW 
\citep{bissantz03}, (ii) the large non-axisymmetric motions of star-forming regions 
\citep{xu06}, (iii)  the moving groups in the SN \citep{dehnen98} containing stars of 
very different ages \citep{famaey05,antoja08}, suggesting that their clumping 
in velocity space is most likely not due to irregular star formation, but rather to 
dynamical perturbations from the bar \citep{dehnen00,minchev10} and/or the spirals 
(e.g., \citealt{qm05}, \citealt{antoja09}), 
or an orbiting satellite galaxy \citep{quillen09,minchev09}.

A numerically well-studied effect of this non-axisymmetry of the Galactic potential is the 
radial migration of stars due to resonant scattering by transient spiral structure (SS)
(\citealt{sellwood02}, hereafter SB02; \citealt{roskar08}), which could explain the 
aforementioned absence of AMR in the SN. 
N-body simulations based on cosmological initial conditions (e.g., \citealt{heller07}) 
would a priori be the most natural framework in which to study 
radial mixing in galactic disks. However, these simulations cannot at present 
produce large enough thin galactic disks due to the transfer of angular momentum from 
the disk to the live dark matter halo. The resulting galaxies are typically thick and
present large stellar velocity dispersions, so that they are closer to S0 galaxies than 
to MW-like ones \citep{sanchez09}. On the other hand, \cite{roskar08} used an 
N-body hydrodynamical model in which the initial conditions were designed to mimic the 
stage following a last major merger when the thin disk formation is supposed to start. 
Both SB02 and \cite{roskar08} compare the effects of radial migration 
in the MW disk, for which the existence of a bar is firmly established, to a simulated 
galaxy lacking a central bar. Unless the effect of the bar is unimportant to the mixing
process, their results would only poorly approximate the radial migration in our Galaxy.
In this paper we check whether the effect of galactic bars can indeed be neglected in the 
context of radial migration in galactic disks, as assumed in a number of previous studies. 
For this purpose we study the combined effect of SS and a central bar (and more generally 
of multiple patterns) on the radial mixing of stars in a stellar disk. 

There is observational evidence for multiple patterns in external galaxies, such as 
asymmetries in the spiral structure (e.g. \citealt{henry03,meidt08}). By expanding galaxy 
images in Fourier components, \cite{elmegreen92} noted that many galaxies exhibit hidden 
three-armed components and suggested that multiple spiral density waves can propagate 
simultaneously in galaxy discs. In addition, by means of the recently developed 
\citep{merrifield06} Radial Tremaine-Weinberg (TWR) method, \cite{meidt09} analyzed the 
high-quality HI and CO data cubes available for four spiral galaxies. The authors found 
direct evidence for the presence of resonant coupling of multiple patterns, including 
spiral-spiral and spiral-bar components. 

The results of N-body simulations provide additional motivation for studying the dynamical 
effects of multiple patterns. \cite{sellwood85} and  \cite{sellwood88} presented N-body 
experiments in which a bar coexists with a spiral pattern moving at a much lower angular 
velocity. \cite{tagger87} and \cite{sygnet88} explained this as a non-linear mode coupling 
between a bar and SS. Sellwood's simulations, and the theoretical explanation of 
\cite{tagger87} and \cite{sygnet88}, were later confirmed by \cite{masset97} and 
\cite{rautiainen99}.

How do multiple patterns affect the dynamics of galactic disks? \cite{quillen03} 
considered the dynamics of stars that are affected by perturbations from both SS and a 
central bar by constructing a one-dimensional Hamiltonian model for the strongest resonances 
in the epicyclic action-angle variables. Quillen pointed out that when two perturbers with
different pattern speeds are present in the disc, the stellar dynamics can be stochastic, 
particularly near resonances associated with one of the patterns. Similar findings were 
presented more recently by \cite{jalali08}. All these results are not surprising since it has 
already been shown by \cite{chirikov79} that in the case of resonance 
overlap the last KAM surface between the two resonances is destroyed, resulting in stochastic 
behavior. We therefore expect that resonance overlap could give rise to both velocity 
dispersion increase and radial migration.

In this paper we study the combined effect of a central bar and SS on the dynamics of a 
galactic disk. We concentrate on a steady-state SS in order to assess the pure effect of 
non-linear coupling between the two perturbers, noting that while there is strong 
circumstantial evidence that SS is mostly not-stationary in unbarred galaxies, the situation 
is less clear in barred ones \citep{bt08}. In any case, the following model can also serve 
as a proxy for studying the effect of transients, and is complementary to the 
transient spiral mechanism developed by SB02.  

\begin{figure}[t!]
\epsscale{1.1}
\plotone{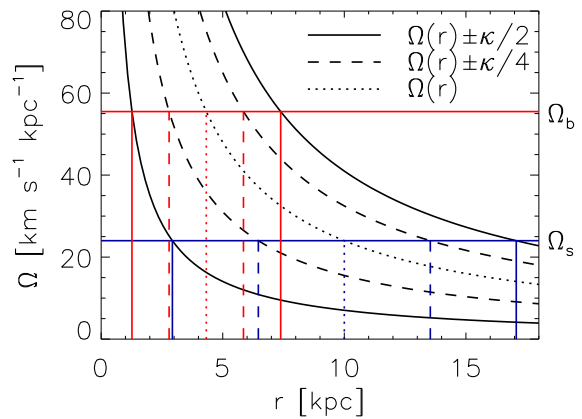}
\figcaption{
Resonances in a galactic disk for nearly circular orbits and a flat rotation curve. Corotation 
occurs along the dotted black curve and is given by $\Omega_{b,s}=\Omega(r)$, where 
$\Omega_{b,s}$ is the bar or spiral pattern speed, $\Omega(r)=v_0/r$ is the local circular 
frequency, and $v_0$ is the constant circular velocity. 
The 2:1 outer and inner Lindblad resonances (OLR and ILR) occur
along the solid black curves, computed as $\Omega_{b,s}=\Omega(r)\pm\kappa/2$, where 
$\kappa$ is the local radial epicyclic frequency. The outer and inner 4:1~LRs occur along 
the dashed black curves, given by $\Omega_{b,s}=\Omega(r)\pm\kappa/4$. The 
red horizontal line indicates the bar pattern speed used in this paper,
$\Omega_b=55.5$~km/s/kpc, and the blue horizontal line shows a spiral pattern speed of 
$\Omega_s=24$~km/s/kpc, such as the one used in the simulations shown in Figs.~\ref{fig:tevol} 
and \ref{fig:eps}. The vertical red and blue lines give the radial 
positions of each resonance for the bar and spiral structure, respectively
(solid lines: 2:1; dashed lines: 4:1, dotted lines: CR). 
\label{fig:om_r}
}
\end{figure}

\section{Resonances in galactic disks}

Galactic disks rotate differentially with nearly flat rotation curves, i.e., constant circular 
velocity as a function of galactic radius. In contrast, density waves, such as a central bar 
or SS, rotate as solid bodies. Therefore stars at different radii would experience different 
forcing due to the non-axisymmetric structure. Of particular interest to us are locations in 
the disk where the stars are in resonance with the perturber. The corotation resonance (CR), 
where stars move with the pattern, occur when the angular rotation rate of stars equals that of 
the perturber. The Lindblad resonances (LRs) occur when the frequency at which a star feels the 
force due to a perturber coincides with the star's epicyclic frequency, $\kappa$. As one 
moves inward or outward from the corotation circle, the relative frequency at which a star 
encounters the perturber increases. There are two values of $r$ for which this
frequency is the same as the radial epicyclic frequency. This is where the inner and outer 
Lindblad resonances (ILR and OLR) are located. Quantitatively, LRs occur when 
$\Omega_{b,s}=\Omega\pm\kappa/m$, where $m$ is the multiplicity of the pattern. The 
negative sign corresponds to the ILR and the positive to the OLR. 
While Bertil Lindblad defined these for the case of an m=2 pattern (thus strictly 
speaking the ILR/OLR are the 2:1 resonances), for an m=4 pattern the ILR/OLR
must be the 4:1 resonances.

Since second order resonances, i.e., 4:1 for a 2-armed spiral or bar, or 2:1 for a 
4-armed spiral, can also be quite important (as will be shown later), we need to have a
convenient way to refer to them. It is somewhat confusing and unclear how the 4:1 
resonances are referred to in the literature. 
The inner 4:1 resonance for an m=2 pattern is known as the Ultra-harmonic resonance 
(UHR). Some also describe the inner and outer 4:1 resonances as the IUHR and the OUHR, others 
as the inner and outer m=4 resonance. If the pattern multiplicity is m=4, then
these become the ILR and OLR. 
To our knowledge, there is no terminology for the 2:1 resonances of an m=4 pattern.
To simplify things, in this paper we generalize the standard notation of Lindblad 
resonances to allow for the existence of both 2:1 and 4:1 resonances, regardless of the 
multiplicity of the pattern. In other words, we will refer to the 2:1~ILR/OLR and the 
4:1~ILR/OLR for both 2-armed (or bar) and 4-armed SS. Naturally, other resonances can also 
be described in this manner, i.e., 3:1, 5:1, 6:1~ILR/OLR, etc.  

Fig.~\ref{fig:om_r} illustrates the relationship between the pattern angular velocity 
and the radii at which resonances occur for a flat rotation curve and nearly circular orbits. 
Note that for two (or more) nonaxisymmetric patterns (such as bar + SS or SS + SS) moving 
at different angular velocities there will always exist regions in the disk where resonances 
overlap. As we will see later, this turns out to be crucial for the dynamics of barred 
galaxies. 

\section{Numerical procedure and choice of parameters}
\label{sec:num}

We follow the motion of test particles in an initially axisymmetric, 2D stellar disk with 
a flat rotation curve as described in detail in \cite{mq07}, in which we grow 
non-axisymmetric perturbations as described below. Both the initial disk stellar density, 
$\Sigma(r)$, and the radial velocity dispersion, $\sigma_r(r)$, decrease 
exponentially with radius, where we initially start with $\sigma_r(r_d)=5$ km/s at the disk 
scale-length, $r_d=3$~kpc for a MW-type galaxy. To simulate the effect of a central bar and 
spiral density waves we impose perturbations on the initially axisymmetric disk as done by 
\cite{mnq07} and \cite{mq07}, respectively. Both bar and spiral perturbations are grown 
simultaneously in 0.4 Gyr by increasing their amplitudes form 0 to the maximum values. 
To ensure a smooth transition to the perturbed state we use the function defined in eq.~4 
from \cite{dehnen00}. To be able to relate parameters to the MW we work in units of the 
solar distance and circular velocity with corresponding values $r_0=8$~kpc and 
$v_c=240$~km/s; this gives $\Omega_0=30$~km/s/kpc. 

\begin{figure*}[t!]
\epsscale{1.1}
\plotone{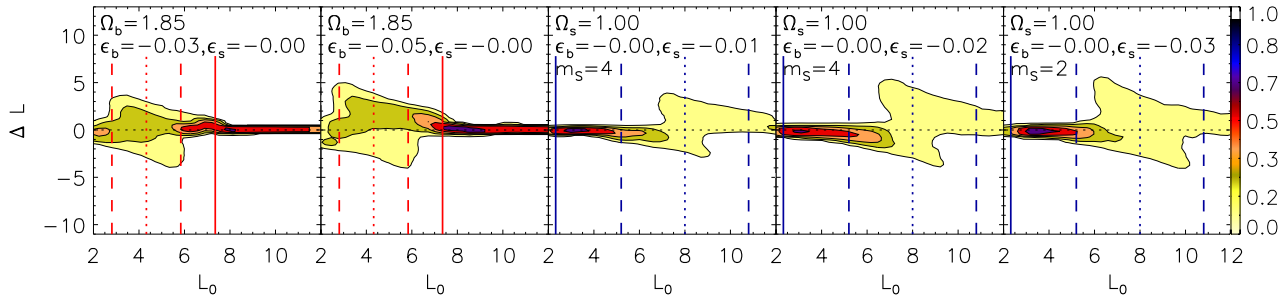}
\figcaption{
Changes in angular momentum, $\Delta L$, as a function of the initial angular momentum, 
$L_0$. Both $\Delta L$ and $L_0$ are in units the circular velocity, $v_c$; $L_0$ is thus 
equivalent to the galactocentric distance, $r$, in kpc. Bar and SS amplitudes are indicated 
in each panel. The first two panels form left to right present simulations with a bar of an 
intermediate strength ($Q_T=0.25$) and a strong one ($Q_T=0.4$), respectively. The third 
and fourth panels show a 4-armed SS with relative overdensity $\Sigma_s/\Sigma_0=0.2,
0.35$. The rightmost panel presents a simulation with a 2-armed SS with an amplitude giving 
rise to $\Sigma_s/\Sigma_0=0.25$. The dotted lines show the corotation radii.
The 2:1 and 4:1~LRs are indicated by the solid and dashed lines respectively (bar=red, 
spiral=blue). Note that, depending on the pattern speed, some resonances might not be present 
in the disk. $|\Delta L|$ increases significantly only near the corotation of
each perturber. 
\label{fig:single}
}
\end{figure*}

\begin{figure*}
\epsscale{1.1}
\plotone{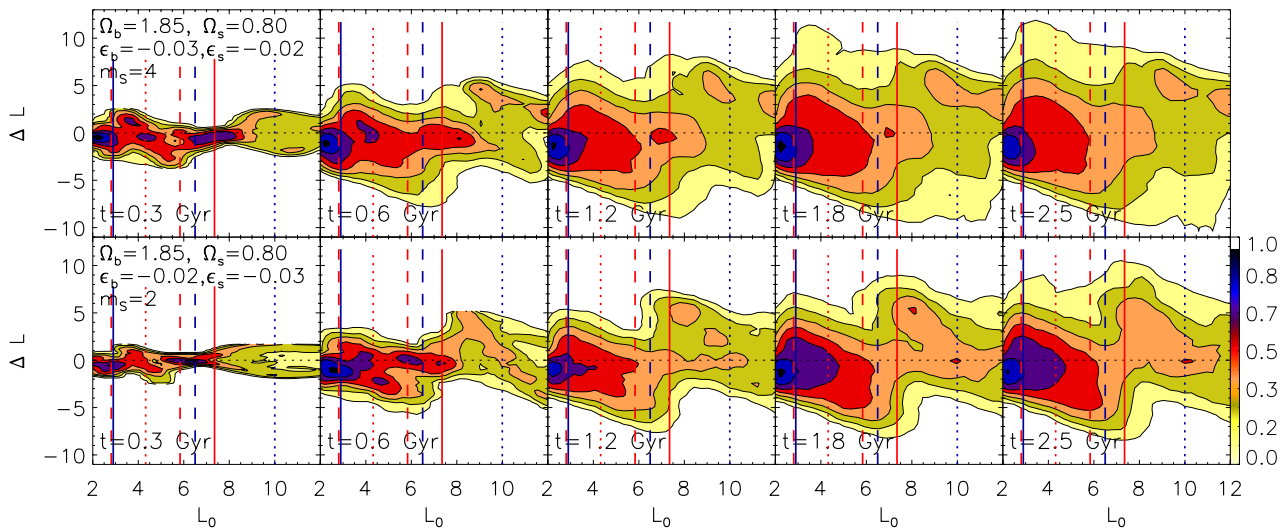}
\figcaption{
Time evolution of the changes in angular momentum, $\Delta L$, for a stellar disk perturbed
by both a central bar and SS. The top row shows a bar and a 4-armed SS of
intermediate strengths: $\epsilon_b=-0.03$ ($Q_T=0.25$) and $\epsilon_s=-0.02$
($\Sigma_s/\Sigma_0=0.35$ for m=4). For the bottom row the bar and the 2-armed SS are weaker:
$\epsilon_b=-0.02$ ($Q_T=0.16$) and $\epsilon_s=-0.03$ ($\Sigma_s/\Sigma_0=0.25$ for m=2).
The increase of $\Delta L$ with time indicates that stars are being placed on radii different 
than their birthplaces, i.e., radial migration takes place throughout the disk. 
\label{fig:tevol}
}
\end{figure*}

While in \cite{mq07} and \cite{mnq07} we required the integration of a large number of 
particles in order to be able to resolve a "solar neighbourhood", here we are interested in 
the global behavior of a galactic disk. Therefore we have found it 
sufficient to integrate $10^5$ particles for each simulation, distributed between an inner 
and outer radii, $r_{in}=2$ and $r_{out}=15$~kpc. We have verified that
increasing the resolution to $10^7$ particles does not change our results.
We divide each sample into 50 radial bins and 
calculate the azimuthally averaged changes in angular momentum in the disk in order to estimate
the induced radial migration. The time evolution of each simulation is followed for 3~Gyr. 
We distinguish between the SS and the bar by indicating the OLR as 
${\rm OLR_s}$ and ${\rm OLR_b}$, respectively, and similarly for the case of the ILR. 

\subsection{Perturbation from a central bar}

The bar strength, $\epsilon_b$, for the case of the MW, is defined as the ratio of the 
forces due to the bar's quadrupole and the axisymmetric background at the Sun's radius, $r_0$, 
on the bar's major axis, $\epsilon_b=3(A_b/v^2_c)(r_b/r_0)^3$ \citep{dehnen00}, where 
$v_c$ is the asymptotic circular velocity. The bar potential is then given by
\begin{equation}
\Phi_{\rm b} = A_{\rm b}(\epsilon_{\rm b}) \cos[2(\phi-\Omega_{\rm b}
t)]\times\left\{
\begin{array}{cclcr}
         \left(r_{\rm b}\over r\right)^3  &,&  r&\ge & r_{\rm b}    \\ 
       2-\left(r\over r_{\rm b}\right)^3  &,&  r&\le & r_{\rm b}
\end{array}
\right.
\label{eq:bar}
\end{equation}
where $r_b=0.8r_{cr}$ is the bar half-length in terms of its CR, $r_{cr}$; in physical 
units $r_b=3.44$~kpc and $r_{cr}=4.3$~kpc.

Previous work has used as a measure of bar strength the parameter $Q_T$ \citep{combes81}. 
This is the ratio of the maximum tangential force to the azimuthally averaged radial force at 
a given radius. From eq.~\ref{eq:bar} this definition yields $Q_T=2A_b/v_c^2$. We examine bar 
amplitudes in the range $0.1<Q_T<0.4$ as expected from observations of various galaxies and 
from N-body simulations \citep{combes81}. This corresponds to $0.013<|\epsilon_b|<0.05$ in 
our units. 

The bar pattern speed is chosen to be $\Omega_b=55.5$ km/s/kpc, which for the case of our 
Galaxy can be written as $\Omega_b=1.85\Omega_0$. This value is consistent with observations 
of external galaxies where it is found that bars are dynamically fast, i.e., 
$r_{cr}/r_b = 1.2\pm0.2$ \citep{aguerri03}, and is also in accordance with recent bar pattern 
speed determinations for the MW \citep{minchev10,mnq07,bissantz03,dehnen00}.

\subsection{Perturbation from spiral structure}
\label{sec:sp}

The spiral potential is given by
\begin{equation}
\Phi_s(r,\phi,t) = \epsilon_s \cos[\alpha \ln{r\over r_0}-m(\phi-\Omega_s t)],
\end{equation}
where $\epsilon_s$ is the SS strength, related to the amplitude of the mass surface
density of spirals, $\Sigma_s$, as 
\begin{equation}
\label{eq:sp}
\epsilon_s \approx - 2 \pi G  \Sigma_s r_0 / (\alpha v_c^2),
\end{equation}
as shown in \cite{bt08}. The parameter $\alpha$ is related to the pitch angle of the spirals, 
$p$, by $\alpha = m \cot(p)$. The azimuthal wavenumber $m$ is an integer corresponding to the 
number of arms. We consider both 2-armed and 4-armed SS with $\alpha=-4$ and -8, 
respectively, where the negative sign corresponds to trailing spirals. \cite{elmegreen98} found 
that grand-design spirals have arm-interarm contrasts of 1.5-6, corresponding to a fractional 
amplitude of $0.2<\Sigma_s/\Sigma<0.7$, which is in agreement with \cite{rix95} who estimated 
$0.15<(\Sigma_s/\Sigma)<0.6$. Since we would like to examine this range of spiral amplitudes 
we need to relate the relative overdensity to the relative potential. 

For a maximum exponential disk the peak circular speed in the disk has the value 
$v_c \simeq 0.622 \sqrt{GM_d / r_d}$ at $r\simeq 2.15r_d$, where $M_d$ is the disk mass inclosed 
by $r_d$. The surface density of the disk at radius $r$ is 
$\Sigma(r) = ({M_d / 2\pi r_d^2}) e^{-r/r_d}$. Eliminating $M_d$ from these expressions, we find
at $r_0$, $v_c^2 \simeq 0.39 \times 2\pi G r_d\Sigma_0 e^{r_0/r_d}$. Substituting this
expression for $v_c$ in eq.~\ref{eq:sp}, the relation between the relative potential 
and the relative overdensity becomes
\begin{equation}
\label{eq:spden}
\epsilon_s \approx - {\Sigma_s\over \Sigma_0} {r_0 \over r_d} {e^{-r_0\over r_d} \\
\over 0.39 \alpha}.
\end{equation}

To correct for finite disk thickness we need to decrease the right-hand side of this equation
by the factor $e^{-kz_0}$ (e.g., \citealt{vandervoort70}), where $k$ is the wave vector given
by $k\approx \alpha/r_0$. For a MW-type galaxy we can adopt a 
disk scale-height of $z_0 \sim300$~pc, resulting in $e^{-kz_0}\approx0.86$ and 0.74 for a 
2- and 4-armed SS, respectively. Using eq.~\ref{eq:spden} and accounting for finite disk 
thickness we estimate that the observed spiral arm amplitudes quoted above would result in 
potential perturbation in the range $0.007<|\epsilon_s|<0.031$ for a 4-armed and
$0.015<|\epsilon_s|<0.072$ for a 2-armed SS.

\begin{figure*}
\epsscale{1.1}
\plotone{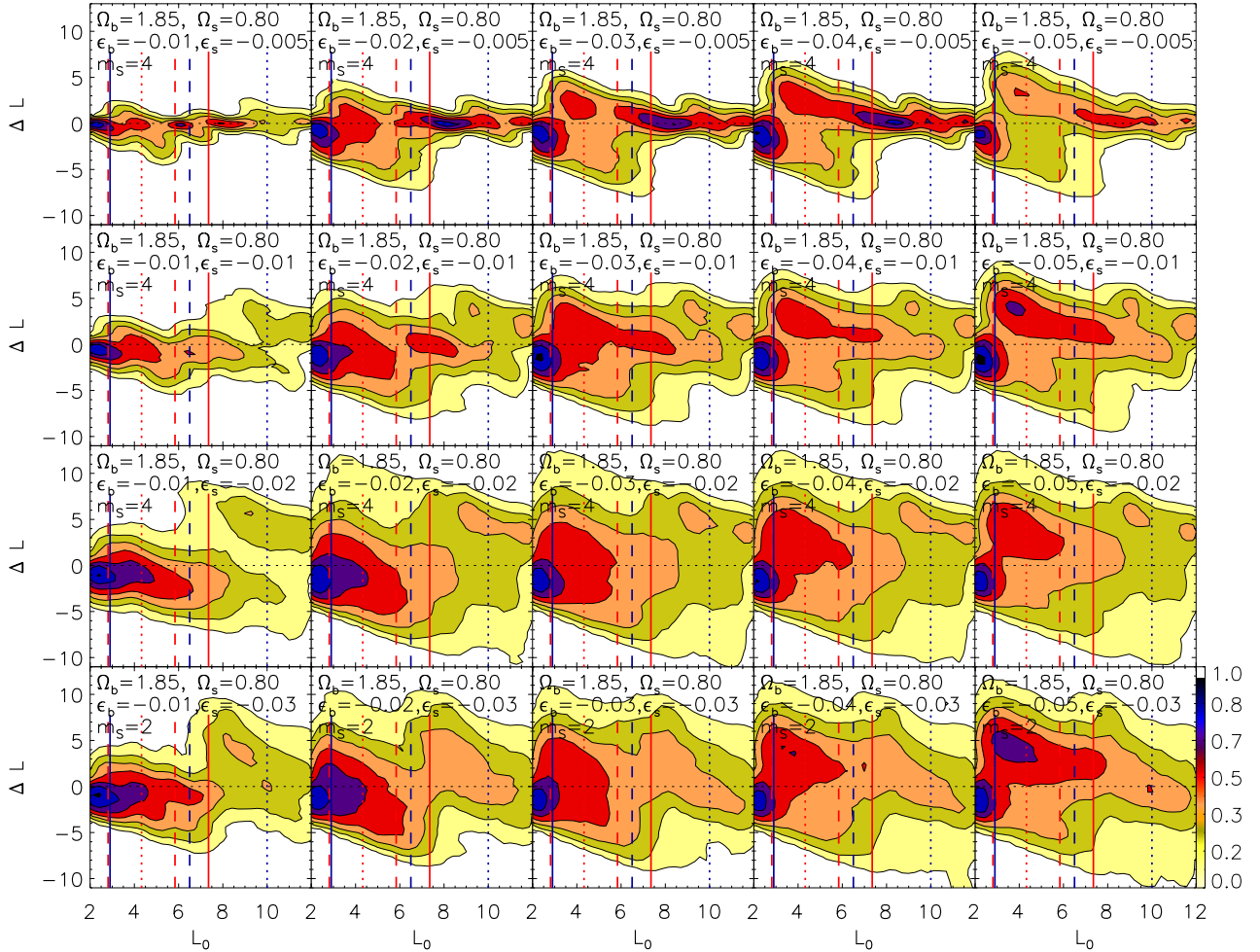}
\figcaption{
The changes in angular momentum at $t=3$~Gyr for simulations with different perturbation 
strengths, while all other parameters are kept fixed. From left to right the bar strength
increases as indicated in each panel. The SS strength increases from top to bottom. Note the 
bimodality caused by the bar's corotation and its 2:1~OLR. 
\label{fig:eps}
}
\end{figure*}

\section{Single perturbation}
\label{sec:single}

We now estimate the changes in angular momentum, $\Delta L$, in a stellar disk. 
To do this, we divide the disk into 50 radial bins and compute $\Delta L=L_1-L_0$, 
where $L_0$ and $L_1$ are the mean values of the initial (at $t=0$) and final (at some later 
time, $t_1$) angular momenta of particles in each bin. 

In Fig.~\ref{fig:single} we show 
$\Delta L$ as a function of $L_0$ for a simulation with a single perturber. Both axes are in 
units of the circular velocity, $v_c$; $L_0$ is thus equivalent to the galactocentric distance, 
$r$ in kpc. The bar and SS amplitudes are indicated in each panel. The first two panels from 
left to right present simulations with a bar of an intermediate strength ($Q_T=0.25$) and a 
strong one ($Q_T=0.4$), respectively. The third and fourth 
panels show a 4-armed SS with relative overdensity $\Sigma_s/\Sigma_0=0.2, 0.35$. The rightmost 
panel presents a simulation with a 2-armed SS with an amplitude giving rise to 
$\Sigma_s/\Sigma_0=0.25$. The dotted lines show the corotation radii. The 2:1 and 
4:1~LRs are indicated by the solid and dashed vertical lines, respectively
(bar=red, spiral=blue). Note that, depending on the pattern speed, certain resonances might not 
be present in the disk. The pattern speeds are given in units of the angular velocity at 
$r_0=8$~kpc, $\Omega_0=30$~km/s/kpc.

Significant angular momentum changes in Fig.~\ref{fig:single} occur only near the CR of each 
perturber (dotted lines). No further increase in $|\Delta L|$ is apparent with time once the 
perturber is fully grown, which is to be expected for a time-independent perturbation. 
We are now interested to see what changes occur when both the bar and SS act together.

\begin{figure*}
\epsscale{1.1}
\plotone{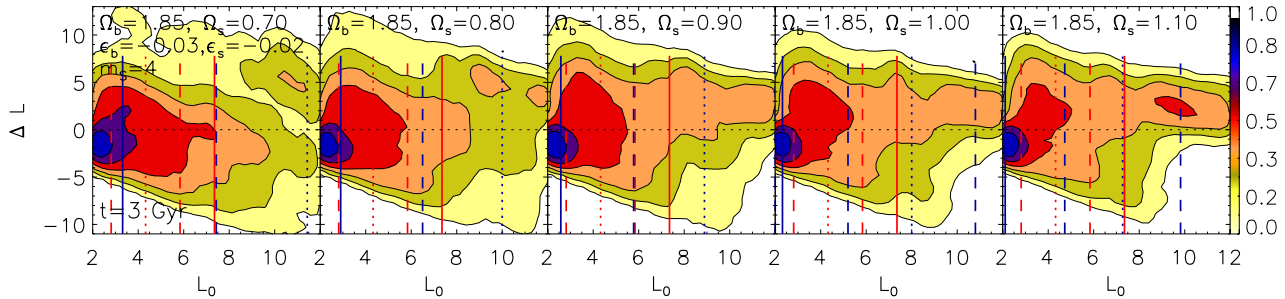}
\figcaption{
Variation of mixing with pattern speed ratio. $\Delta L$ is shown for different SS 
pattern speeds, $\Omega_s$, while keeping $\Omega_b$ the same. This results in different 
spacing of the bar and SS resonances. From left to right the 4-armed SS moves at 
$\Omega_s=0.7, 0.8, 0.9, 1.0,$ and 1.1 in units of $\Omega_0=30$~km/s/kpc. All other 
simulation parameters are kept fixed: the time is $t=3$~Gyr and the strengths of the 
perturbers are the same as in the top row of Fig.~\ref{fig:tevol}. As expected, the effect 
of resonance overlap on $\Delta L$ is strongest when the highest order resonances 
(2:1~$\rm OLR_b$ and 4:1~$\rm ILR_s$) lie closest to each other (leftmost panel, $L_0\sim7.5$).
\label{fig:om}
}
\end{figure*}

\section{Bar + Spiral Structure}
\label{sec:spbar}

As shown by \cite{mq06}, when two perturbers moving at different pattern speeds are imposed on 
a galactic disk, an increase in the random motions of stars is expected, i.e., the disk heats. 
In addition, the authors found that in regions of resonance overlap stars drift radially with 
time from their birth radii. This suggests that such a mechanism could be responsible for 
radial migration in the disk. Here we quantify this idea more clearly for the case of bar 
+ SS, by examining the effect of the simultaneous propagation of the two perturbers.

\subsection{Time evolution of $\Delta L$}
\label{sec:tevol}

In Fig.~\ref{fig:tevol} we show the time development of the changes of angular momentum in a 
stellar disk perturbed by both a central bar and SS. In each row we present a different 
combination
of bar and SS strengths, as described bellow, while all other parameters are kept fixed. As in 
Fig.~\ref{fig:single}, contour plots show the change in angular momentum, $\Delta L$, as a 
function of the initial angular momentum $L_0$ (or radius, $r$ in kpc). Panels from left to 
right show the temporal evolution of the system for 0.3, 0.6, 1.2, 1.8 and 2.5 Gyr. 
In the top row both the bar and the 4-armed SS have intermediate
strengths: $\epsilon_b=-0.03$ ($Q_T=0.25$) and $\epsilon_s=-0.02$ ($\Sigma_s/\Sigma_0=0.35$ 
for m=4). For the bottom row the bar and the 2-armed SS are weaker:
$\epsilon_b=-0.02$ ($Q_T=0.16$) and $\epsilon_s=-0.03$ ($\Sigma_s/\Sigma_0=0.25$ for m=2). 
The pattern speeds in both simulations are the same so that we can asses the effect of the 
different amplitudes.

Examining the first row of Fig.~\ref{fig:tevol}, we note that at the beginning of the simulation 
$|\Delta L|$ increases mainly at the CR of each perturber (dotted lines), which would be the 
case of linearly adding the individual effects of the bar and SS seen in Fig.~\ref{fig:single}. 
However, at later times large changes in angular momentum occur, suggesting that non-linear 
effects become important. Note that at $t=0.6$~Gyr the effect near each CR is already stronger 
than that expected from each individual perturber (compare to Fig.~\ref{fig:single}).
Meanwhile the area between the CRs of the bar and SS fills in
with time as a result of the proximity of the 4:1~$\rm ILR_s$, the 4:1~${\rm OLR_b}$ and the 
2:1~${\rm OLR_b}$, thus extending the effect over the entire disc. At $t=1.8$~Gyr (fourth panel 
of Fig.~\ref{fig:tevol}), the changes in angular momentum throughout the disk are similar in
amplitude to those resulting from recurrent spirals (Fig.~11 in SB02). To reach this level of 
mixing, however, requires less than $\sim2$~Gyr compared to $\sim9$~Gyr for the case of 
transients. 

Decreasing the bar strength to $\epsilon_b=-0.02$ ($Q_T=0.16$) (bottom row of 
Fig.~\ref{fig:tevol}) and considering a slightly weaker 2-armed SS has a weaker effect on 
$\Delta L$. However, changes in $\Delta L$ are still very strong, mixing the disk in less than 
3~Gyr. 

At this time we can already identify two main differences between radial migration
from transients and the mechanism described in this work: (1) The time-scale for 
disk mixing is much shorter in the case of resonance overlap and (2) resonance overlap causes 
a bimodal distribution in $\Delta L$ with peaks near the bar's CR and its 2:1~OLR, compared to 
the single peak seen in Fig.~11 and 12 in SB02. 
Next we explore further the effect of perturbation strengths.

\subsection{Variation with perturbation strengths}
\label{sec:strength}

The strong increase of $|\Delta L|$ with time when both a bar and SS perturb a stellar disk 
suggests that this is a nonlinear effect. We therefore expect strong variation with the strength 
of the perturbers. To investigate this further we now explore the effect on mixing when different 
combinations of $\epsilon_s$ and $\epsilon_b$ are considered. In Fig.~\ref{fig:eps} we plot the 
changes in angular momentum at $t=3$~Gyr for simulations with different perturbation strengths, 
while all other parameters are kept fixed. From left to right the bar strength increases as 
indicated in each panel. The SS strength increases from top to bottom.

Observing the top, leftmost panel of Fig.~\ref{fig:eps} we see that when both perturbers are
weak mixing is not very strong, although an increase in $|\Delta L|$ near the 2:1~$\rm OLR_b$ 
is still apparent. 
However, as the bar strength increases, while keeping the spiral weak, the resonance overlap in 
the inner disk has a surprisingly strong effect on $\Delta L$. To realize this compare the top 
row of Fig.~\ref{fig:eps} to the corresponding bar strengths in Fig.~\ref{fig:single}. Similar 
behavior is observed in the case of a weak bar and a range of SS strengths (first column in 
Fig.~\ref{fig:eps}). Note that here in addition to the inner disk we also see a strong effect 
near the spirals' CR at 10~kpc, which is well outside the 2:1~$\rm OLR_b$. A possible 
explanation could be the proximity of the CR of SS to the bar's 1:1 asymmetric, 1:1 symmetric 
resonances, or $x_1^*(2)$ unstable orbits \citep{dehnen00}.

\subsection{Variation with pattern speed ratio}
\label{sec:ratio}

In Fig.~\ref{fig:om} we plot $\Delta L$ for different SS pattern speeds, $\Omega_s$, while 
keeping $\Omega_b$ the same. This results in different spacing of the bar and SS resonances. 
From left to right the 4-armed SS moves at $\Omega_s=0.7, 0.8, 0.9, 1.0,$ and 1.1 in units of 
$\Omega_0=30$~km/s/kpc. All other simulation parameters are kept fixed: the time is $t=3$~Gyr 
and the strengths of the perturbers are the same as in the top row of Fig.~\ref{fig:tevol}. As 
one would expect, the effect of resonance overlap on $\Delta L$ is strongest when the highest 
order resonances (2:1 $\rm OLR_b$ and 4:1 $\rm ILR_s$) lie very close to each other (leftmost 
panel, $L_0\sim7.5$). For all pattern speed ratios the distribution of $\Delta L$ exhibits
a bimodality near the bar's CR and near its 2:1 OLR. This signature can be used
to identify the mechanism described here in fully self-consistent simulations, where the
disk dynamics is much more complicated and may include additional mixing due to transient 
spirals and small mergers. 

\section{Implications for the Milky Way}
\label{sec:mw}

We showed in Sec.~\ref{sec:spbar} that the simultaneous propagation of a central bar and SS 
in a stellar disk induces radial migration. Since this is a nonlinear effect, depending on the 
perturbers' amplitudes, strong mixing could occur in a short period of time. Similarly to the 
majority of disk galaxies, the MW has been well established to be a barred galaxy, thus 
we expect mixing due to resonance overlap to be at work here as well. While there is no doubt 
that the MW has been affected by this process, it may be difficult to estimate exactly 
how. This migration mechanism is a strong function of the strengths of the MW bar and 
SS. However, it may be incorrect to use the currently observed spiral and bar amplitudes since
most likely they have not been the same throughout the lifetime of the Galaxy. In fact, we 
expect that spirals were stronger in the past, when the disk velocity dispersion was lower, thus 
resulting in a more gravitationally unstable disk. In any case, given that the past history of 
the MW spiral and bar structure is largely unknown, we will make an estimate of how much mixing 
could have resulted by using the currently observed SS and bar parameters.

\subsection{Choice of bar and spiral structure parameters}

For all simulations in this paper we have used a bar pattern speed of $\Omega_b=1.85\Omega_0$, 
where $\Omega_0$ is the angular velocity at the solar circle $r_0$. For $r_0=8$~kpc and 
$v_c=240$~km/s, in physical units $\Omega_b=55.5$~km/s/kpc, consistent 
with recent pattern speed estimates \citep{minchev10,mnq07,bissantz03,dehnen00}.
\cite{rodriguez08} estimated the strength of the MW bar to lie in the range 
$0.25<Q_T<0.4$ by fitting bar models to near infrared observations (2MASS counts).
In the units used in this paper this range corresponds to $0.03<|\epsilon_b|<0.05$.
The individual effect on $\Delta L$ of bars with amplitudes $\epsilon_b=-0.03$ and
$\epsilon_b=-0.05$ can be seen in Fig.~\ref{fig:single}.

\begin{figure*}
\epsscale{1.1}
\plotone{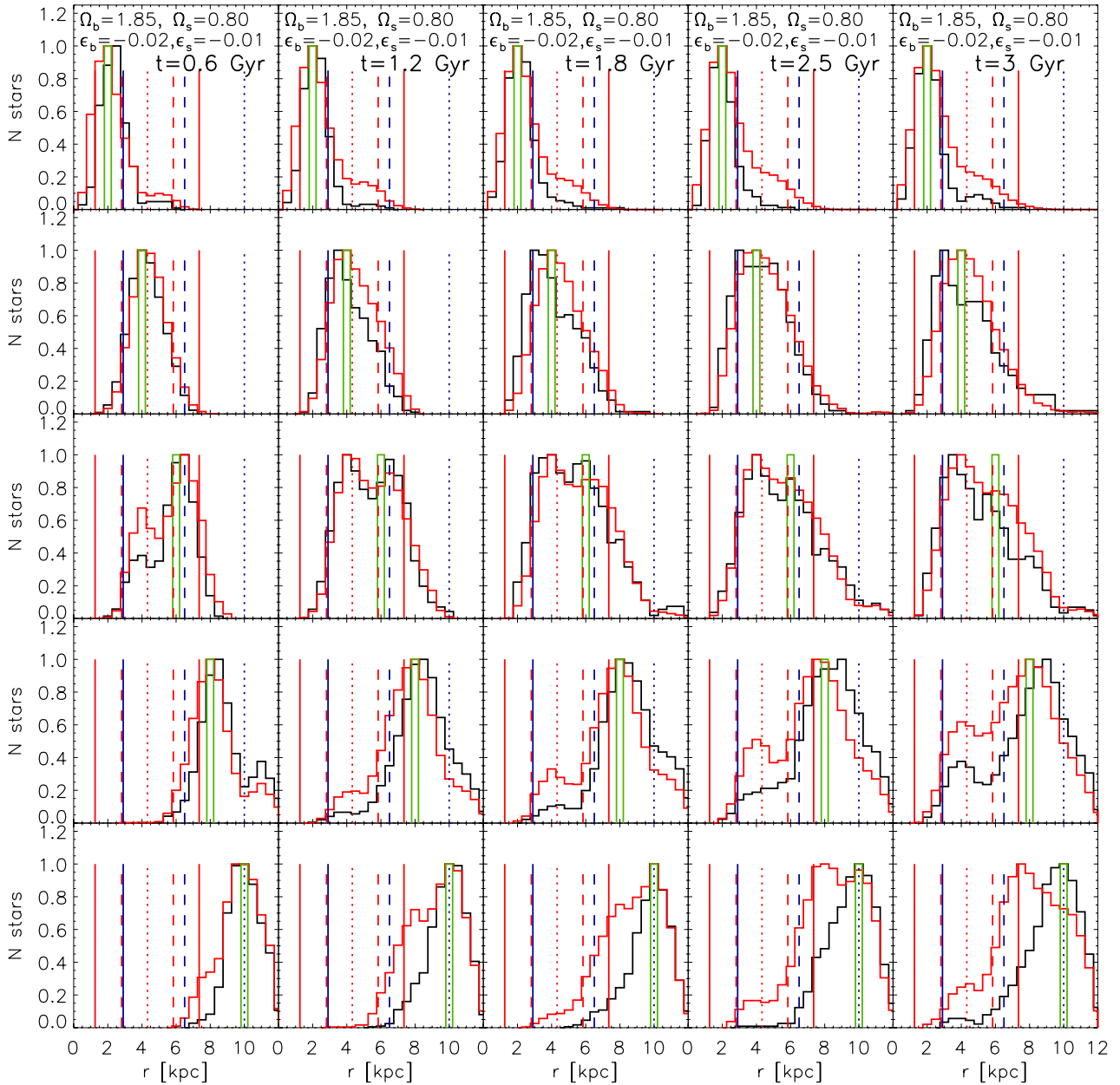}
\figcaption{
The distribution of the initial radii of stars, which have ended up in an annulus of 400 pc 
(in the green bin) centered at a given galactocentric distance, $r$. Simulation parameters
are consistent with the Milky Way structure (see Sec.~\ref{sec:mw}). From top to bottom 
different rows show the time evolution of each such annulus centered on $r=$~2, 4, 6, 8, and 
10~kpc. The black and red histograms show stars on nearly
circular orbits ($\sigma_r\sim5$ km/s) and the total population, respectively. With this 
new mechanism, the solar neighborhood stars could have come from a wide range of Galactic 
radii, including from the outer Galaxy, and thus account for the flatness of the observed AMR.
Note the different signature in the mixing of cold and hot orbits.
\label{fig:rhist}
}
\end{figure*}

\begin{figure}[t!]
\epsscale{1.0}
\plotone{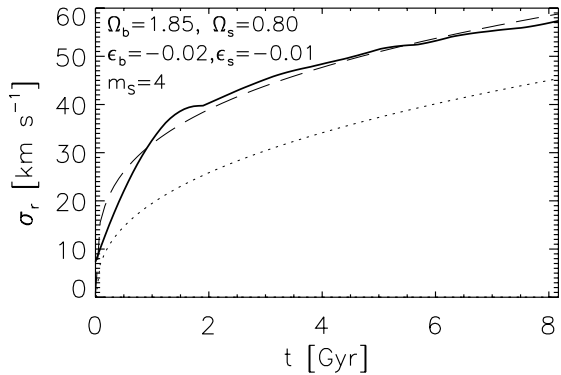}
\figcaption{
Radial velocity dispersion as a function of time at the solar radius (solid line) for the
simulation in Fig.~\ref{fig:rhist}. The dashed line shows a power-law fit to the simulated data. 
The dotted line plots the observed age-velocity relation in the solar neighborhood 
\citep{holmberg09}.
\label{fig:sig}
}
\end{figure}
The MW SS parameters are much more uncertain. It has been proposed to be a superposition of 2- 
and 4-armed patterns moving near corotation with the Sun \citep{lepine01,torra00}. On the other 
hand, \cite{martos04} and \cite{bissantz03} matched the properties of the gas in the inner Galaxy 
with a 4-armed SS with a pattern speed of 20~km/s/kpc (or with the Sun near the 4:1~ILR). 
The MW SS at this stage in the Galaxy evolution appears to be quite weak compared to 
observations of external galaxies: \cite{drimmel01} derived $\Sigma_s/\Sigma_0=0.15$ for the 
stellar component (based on K-band observations). These authors, however, suggested that most 
likely this value is larger. There are also indications from
the gas flow in the inner Galaxy that the MW has stronger spiral arms, where the amplitude 
in the mass density is larger by a factor of 1.5 than its amplitude in the near-infrared 
luminosity density (see \citealt{bissantz03,fb05}). Therefore, a possible range for the relative 
overdensity is $0.15<\Sigma_s/\Sigma_0<0.23$. These values, including the correction for finite 
disk thickness (Sec.~\ref{sec:sp}), correspond to a potential perturbation of 
$0.015<|\epsilon_s|<0.024$ and $0.007<|\epsilon_s|<0.01$ for a 2- and 4-armed SS, respectively. 

From the discussion above, for parameters consistent with our Galaxy, the amount and rate of 
mixing due to spiral-bar resonance overlap is given by the second row of Fig.~\ref{fig:eps} 
for a 4-armed SS, and by the bottom row for a 2-armed SS (excluding the leftmost panels in 
both cases). Note that the strength of the 2-armed SS in Fig.~\ref{fig:eps} is slightly 
stronger than the estimated range.

\subsection{Mixing at different Galactic radii}

To see in greater detail how stars migrate throughout the disk we now examine the effect on 
mixing at different galactic radii. To be consistent with MW parameters, we use the simulation 
for which we presented the changes in angular momentum in Fig.~\ref{fig:eps}, second row, second 
column. In Fig.~\ref{fig:rhist} we show the distribution of birth radii (at $t=0$) of stars 
which are "currently" (at the time indicated in each plot) in an annulus of 400~pc (in the green 
bin). Each row shows an annulus centered on a different galactic radius, $r$: from top to bottom 
$r=2$, 4, 6, 8, and 10~kpc. To show the effect on stars placed on nearly circular orbits, the 
black histograms plot only stars with a cut in the eccentricity, $e<12$. We estimate $e$ as
$e\approx\sqrt{(u^2+2v^2)/2}$, appropriate for a flat rotation curve \citep{arifyanto06}, and
$u,v$ are the radial and tangential velocity components of stars with the circular velocity 
subtracted. The radial velocity dispersion resulting for this sample is $\sim5$~km/s, insuring 
that stellar orbits are almost circular. The red histograms show the full sample. Stars on cold 
orbits, shown by the (normalized) black histograms, constitute about 10\% of the total 
population. 

Firstly, note that the distributions peak at radii near regions of resonance overlap, as 
expected. While in the inner disk ($r=2$ and 4~kpc) there are strong weights at larger radii, 
the situation is reversed for $r\ge6$~kpc. At the solar distance, $r=8$~kpc (fourth row in 
Fig.~\ref{fig:rhist}), a large population of stars from the bar's CR ($r\sim4.5$~kpc)
enters the solar circle. Naturally, when the total population is considered (red histogram) more 
stars enter from the inner disk due to the exponential profiles of both the density and 
the radial velocity dispersion. Observe also the overall shift toward larger radii of the cold 
population compared to the total sample. 

While the pattern speed of the MW SS is not well constrained, the bar is well established to 
have its 2:1 OLR just inside the solar circle. Since the bar's CR and OLR are both strong and 
roughly invariable (only small changes in the bar parameters are expected in $\sim3$~Gyr), we 
can expect to find a bimodal distribution of $\Delta L$ with peaks at these bar resonances 
{\it regardless of the SS pattern speed}. That is exactly what is seen in Fig.~\ref{fig:om}. 
What if the MW spirals were transient? In that case the distribution of the changes in angular 
momentum can be approximated by the linear combination of a range of simulations with different 
SS pattern speeds (imagine adding up all panels in Fig.~\ref{fig:om}). This would again result 
in a bimodal distribution. Therefore, although the past history of the MW structure is not 
known and a large range of possible perturber strengths, pattern speeds and longevity exists, 
{\it it would be reasonable to expect to always find a bimodality in the distribution of 
$\Delta L$}. 

Consequently, the large fraction of stars coming from the bar's CR which enters the solar circle 
(fourth row of Fig.~\ref{fig:rhist}) may be one of the observational signatures of the spiral-bar 
resonance overlap mechanism. This is expected to affect the local metallicity distribution of 
stars. We also note that a large fraction of stars comes from the outer disk, roughly 40\% of 
all stars for this particular pattern speed ratio. This is consistent with the metal-poor stars 
from the outer Galaxy scattered into the SN, as recently inferred from observational data 
\citep{haywood08}, and a metal-rich population coming from the inner Galaxy. With this new 
mechanism, the SN stars could have come from a wide range of Galactic disk radii, including from 
the outer Galaxy, and thus can account for the flatness of the observed AMR.
 
On the other hand, away from the solar circle we expect to find overdensities of stars 
originating at the bar's 2:1 OLR. We clearly see this in the initial distribution of stars ending 
up in the annuli centered on $r=2, 4, 6$ and 10~kpc (Fig.~\ref{fig:rhist}). 
This may provide another observable for identifying the currently discussed mechanism.

However, the most promising technique to put constraints on this mechanism in the MW is 
"chemical tagging" \citep{freeman02}. 
The idea is to use the detailed chemical abundance patterns of 
individual stars in our Galaxy to associate them with common ancient star-forming aggregates 
with similar abundance patterns. This technique will be very important to 
characterize the star formation and chemical histories of the MW. Chemical tagging will become
possible with the HERMES survey on the Anglo-Australian Telescope (AAT; 
\citealt{freeman08,freeman10}) and the APOGEE survey at the Apache Point Observatory 
\citep{allende08}. Therefore, in the next couple of years we may be in a position to 
directly measure the spread of open clusters across the Galaxy as a function of age, thus 
providing direct constraints on the amount of mixing in the Milky Way \citep{joss10}.

\subsection{The effect on disk heating}
\label{sec:sig}

Now we look at the effect on the increase of random motion of stars with time. The observed 
correlation between the ages and velocity dispersions of SN stars has been a subject of study 
since the work of \cite{spitzer51, spitzer53} (see \citealt{mq06} for a detailed introduction). 
Deferent mechanisms have been proposed to explain the age-velocity relation (AVR) (or stellar 
disk heating), including scattering by giant molecular clouds (GMC)
(e.g., \citealt{mihalas81}) and transient spirals (e.g., \citealt{barbanis67}). In \cite{mq06} 
we showed that the resonance overlap of multiple spiral patterns provides yet another mechanism 
for disk heating. We therefore expect that an overlap of bar and SS resonances would give a 
similar result. To check if this is indeed the case, in Fig.~\ref{fig:sig} we plot the radial 
velocity dispersion, $\sigma_r$, (solid line) as a function of time for the same simulation used
to produce Fig.~\ref{fig:rhist}. The dashed line shows a power-law fit to our simulated data: 
$\sigma_r\sim t^{0.29}$. To compare to observations we also show the AVR in the SN (dotted line), 
as estimated recently from analysis of the Geneva-Copenhagen Survey data \citep{holmberg09}. 
Note that the observed AVR goes like $\sigma_r\sim t^{0.39}$. Moreover, in our simulation the 
disk heats $\sim20\%$ more than observed. 
There could be several possibilities that could account for this
discrepancy: (1) our simple model does not include a gaseous disk component nor star formation, 
which would naturally lower the velocity dispersion; (2) while here we considered 2D disks,
a significant fraction of the heating in the galactic disk plane would be transferred 
to the vertical direction, where GMC serve as scattering agents \citep{jenkins90,jenkins92}; 
and (3) in this work we used constant parameters for both the MW bar and SS, but it is 
reasonable to expect more realistic scenarios, where throughout the lifetime of the Galaxy 
there could have been variations of the perturbers' amplitudes and pattern speeds, bar size, 
spiral pitch angle, as well as bar destruction and reformation \citep{bournaud02}.

\section{Discussion and conclusions}

In this work we have examined the combined effect of a central bar and spiral structure on the 
radial migration in galactic disks. We have found that:

\noindent (i) This new mechanism for galactic disk mixing could be up to an order of magnitude 
more effective than the transient spirals one (see Sec.~\ref{sec:strength}). We note that the 
effect is non-linear, strongly dependent on the strengths of the perturbers. The non-linearity 
of the mechanism is confirmed by the fact that an increase in the amplitudes yields a much 
larger increase in the angular momentum changes in a simulation including {\it both} bar 
and spiral structure, compared to the case of a single perturber. In other words, the individual 
effects do not add up linearly. Another indication of non-linearity is that the effect increases 
with time only in the case of the two perturbers propagating simultaneously. 

\noindent (ii) The signature of this mechanism is a bimodality in the changes of angular
momentum in the disk with maxima near the bar's corotation and its outer Lindblad resonance 
(see Figs.~\ref{fig:om} and \ref{fig:rhist}). This is true regardless of the spiral pattern 
speed and can be used in identifying the spiral-bar resonance overlap migration in 
N-body simulations, where the disk dynamics is much more complicated.

\noindent (iii) For bar and spiral parameters consistent with Milky Way observations we find 
that it takes $\sim$~3~Gyr to achieve the mixing for which transients require 9~Gyr. 
In addition to radial 
mixing, spiral-bar coupling can account for the age-velocity relation (AVR) observed in the 
solar neighborhood (Fig.~\ref{fig:sig}, Sec.~\ref{sec:sig}). Note however that the estimates 
for the Milky Way spiral and bar strengths derived from observations {\it now} may simply be 
irrelevant for our understanding of the current state of mixing in the Galactic disk 
(see Sec.~\ref{sec:mw}).

\noindent (iv) This mechanism could explain observations showing that strongly barred 
galaxies have weaker metallicity gradients than weakly barred or non-barred galaxies
(e.g., \citealt{zaritsky94,martin94,perez09}).

The effect of resonance overlap we have described in this paper appears to be quite strong. 
In order for it to work, a galactic disk must contain a central bar\footnote{Note that 
resonance overlap, and thus migration, will also result from multiple spiral structure 
\citep{mig2}.}. Unlike in our test-particle simulations, where the disk responds solely to 
the bar potential, as seen in N-body simulations and deduced from observations, a bar is 
expected to always drive spiral structure (e.g., \citealt{salo10}). 
Thus, any barred galaxy would be affected strongly by this mechanism. As the disk heats and SS 
diminishes in strength, the effect on $\Delta L$ will decrease. It is therefore likely that
resonance overlap migration will be most vigorous at certain periods during a galaxy lifetime. 

In this work we have only considered steady-state spirals in order to assess the pure effect of 
the coupling between bar and spiral structure. Considering transient spirals would increase the 
efficiency, since the SB02 mechanism would come into play as well. Therefore, a combination of a 
central bar and short-lived transients is expected to be extremely efficient at mixing barred 
spiral galaxies. 

As discussed in Sec.~\ref{sec:mw}, the most promising technique to put constraints on 
this mechanism in the MW is "chemical tagging" \citep{freeman02, joss10} which will become 
possible with the forthcoming spectroscopic survey HERMES \citep{freeman08,freeman10}. More 
ongoing and planned large Galactic surveys, such as RAVE, SEGUE, SIM Lite, GYES, LAMOST and 
GAIA, can search for signatures of the mechanism (see also \citealt{mq08}). 

In a follow-up paper \citep{mig2} we study the effect of resonance overlap in fully 
self-consistent, TreeSPH/N-body simulations including a gaseous component and star formation.
The migration mechanism described in this paper can provide an explanation for the observed 
flatness and spread in the age-metallicity relation. Our results, however, need to be 
incorporated into a chemo-dynamical galactic model to properly assess the effect on the AMR,
and possibly also on the formation of the thick disk, similarly to the work by 
\cite{schonrich09}.

\acknowledgements
We would like to thank Bruce Elmegreen, Ken Freeman, Arnaud Siebert, Jerry Sellwood, and 
James Binney for helpful communications and discussions. We also thank the anonymous referee 
for valuable suggestion which have greatly improved the manuscript. Support for this work was 
provided by ANR and RAVE. 

\bibliography{myreferences}
\bibliographystyle{astron}

\end{document}